\documentclass[twocolumn,aps,showpacs,floatfix,prc]{revtex4}
\usepackage[dvips]{epsfig}
\begin{document}

\title{$J/\psi$  production  in  Au+Au collisions at RHIC and the
nuclear absorption}

\author{A. K. Chaudhuri}
\email[E-mail:]{akc@veccal.ernet.in}
\affiliation{Variable Energy Cyclotron Centre,\\ 1/AF, Bidhan Nagar,
Kolkata 700~064, India}

\begin{abstract}
It  is  shown that a QCD based nuclear absorption model, with few
parameters fixed to reproduce  experimental $J/\psi$  yield  in  200  GeV
pp/pA   and   450  GeV  pA  collisions  can  explain  the  preliminary 
PHENIX data  on  the  centrality  dependence  of   $J/\psi$
suppression    in    Cu+Cu    collisions    at    RHIC    energy,
$\sqrt{s_{NN}}$=200 GeV. However, the model does not give
satisfactory description to the preliminary PHENIX data on the
centrality
dependence  of $J/\psi$ suppression  in Au+Au
collisions. The analysis suggest that in Au+Au collisions, $J/\psi$ are
suppressed in a medium unlike the medium produced in
SPS energy nuclear collisions or in RHIC energy Cu+Cu collisions.
\end{abstract} 
\pacs{PACS numbers: 25.75.-q, 25.75.Dw}

\maketitle

\section{introduction}

Lattice  QCD predicts that under certain conditions (sufficiently
high energy density and temperature),  ordinary  hadronic  matter
(where  quarks  and  gluons  are  confined),  can undergo a phase
transition to a deconfined matter, commonly known as Quark  Gluon
Plasma (QGP). Nuclear physicists are trying to produce and detect
this  new  phase  of matter at RHIC, BNL. $J/\psi$ suppression is
recognized as one of the promising signal  of  the  deconfinement
phase  transition.  Due to screening of color force, binding of a
$c\bar{c}$ pair into a $J/\psi$ meson will be  hindered,  leading
to  the  so  called  $J/\psi$ suppression in heavy ion collisions
\cite{Matsui:1986dk}.  However,  $J/\psi$'s  are   absorbed   in   nuclear
collisions  also  and  prior  to  NA50  158 AGeV Pb+Pb collisions
\cite{Abreu:2000ni},  all  the   experimental   data   on   $J/\psi$
suppression,  are  explained  with  nuclear absorption only. NA50
collaboration  measured   centrality   dependence   of   $J/\psi$
suppression  in  158  AGeV  Pb+Pb collisions. Data gave the first
indication of 'anomalous'  mechanism  of  charmonium  suppression
which  goes  beyond the conventional nuclear absorption. The data
generated lot of excitement as it  was  believed  to  give  first
indication  of QGP formation. Later, the data were explained in a
variety  of  models,  with  or  without  the  assumption  of  QGP
\cite{Capella:2000zp,Kostyuk:2003kt,Gorenstein:2000ck,
Grandchamp:2003uw,Blaizot:2000ev,Chaudhuri:2001af,
Chaudhuri:2001zx,Chaudhuri:2003zs,Chaudhuri:2002uf,
Chaudhuri:2006zg,Qiu:1998rz}.       More
recently, NA60 collaboration measured the  centrality  dependence
of   charmonium   suppression   in   158  AGeV  In+In  collisions
\cite{Shahoyan:2005jj,Borer:2005qy,Arnaldi:2006ee}. In  In+In  collisions   also,   one   observes
anomalous  suppression,  which is beyond the conventional nuclear
absorption.

In  recent  Au+Au  collisions  at  RHIC,  one  observe a dramatic
suppression of hadrons with high  momentum,  transverse  to  beam
direction            (high           $p_T$           suppression)
\cite{BRAHMSwhitepaper,PHOBOSwhitepaper,PHENIXwhitepaper,STARwhitepaper}.
This has been interpreted as evidence for the  creation  of  high
density,  color  opaque  medium  of  deconfined quarks and gluons
\cite{QGP3jetqu}. It is expected that high density, color opaque medium
will leave its imprint on $J/\psi$ production. At RHIC energy, it has been
argued that rather than suppression, charmoniums will be enhanced
\cite{ Thews:2000rj,Braun-Munzinger:2000px}. 
Due to large initial energy, large number of $c\bar{c}$ pairs will be
produced in initial hard scatterings. Recombination of $c\bar{c}$
can occur enhancing the charmonium production.
PHENIX  collaboration  have measured the centrality dependence of
$J/\psi$ invariant yield in  Au+Au  collisions  at  RHIC  energy,
$\sqrt{s_{NN}}$=200  GeV \cite{Adler:2003rc,Nagle:2002ib}.  More recently, with improved statistics, they have
measured $J/\psi$'s in Cu+Cu and in Au+Au collisions.
  Preliminary
results for the centrality dependence of nuclear modification factor ($R_{AA}$) and mean square transverse momentum for $J/\psi$ suppression in Cu+Cu and in Au+Au collisions are available
\cite{Nagle:2005sh,PereiraDaCosta:2005xz}.
 PHENIX data on
$J/\psi$ production in Au+Au/Cu+Cu collisions, are not consistent
with models which predict $J/\psi$ enhancement
\cite{ Thews:2000rj,Braun-Munzinger:2000px}. It was also  seen
that  various  models,  e.g.  comover  model   
\cite{Capella:2000zp}, statistical  coalescence  model    
\cite{Kostyuk:2003kt}  or the kinetic model  \cite{Gorenstein:2000ck}
fail to explain the PHENIX  (preliminary)  data  on  the  nuclear
modification   factor   for   $J/\psi$  in  Cu+Cu  and  in  Au+Au
collisions. The data are also not  explained  in  normal  nuclear
absorption model \cite{Vogt:2005ia}.

We have developed a QCD based nuclear absorption model to explain
the  anomalous $J/\psi$ suppression  in 158
AGeV  Pb+Pb  collisions  \cite{Chaudhuri:2001zx,Qiu:1998rz}. 
Unlike in conventional nuclear absorption model, in the QCD based
nuclear absorption model, the $c\bar{c}$ pair interact with the medium and gain relative 4-square momentum. 
Some of the pairs can gain enough 4-square momentum to cross the
threshold for open charm meson, reducing the $J/\psi$ yield.
The parameters of the
model were fixed to reproduce $J/\psi$ yield in  pp  and  pA  collisions.
The model give consistent  description  of the
centrality dependence of the $J/\psi$ suppression and $p_T$ broadening 
in 158
AGeV Pb+Pb  collisions    and  in  200  AGeV  S+U
collisions \cite{Chaudhuri:2002uf}. 
In  the  present  paper  we have
tested  the  model  against  
the  preliminary PHENIX data on $J/\psi$ suppression in
Cu+Cu and Au+Au collisions at  RHIC  energy,  $\sqrt{s_{NN}}$=200
GeV. Centrality dependence of $J/\psi$ suppression, in Cu+Cu collisions, is well explained in the model, but 
the model fails to explain the suppression in
Au+Au collisions.  The analysis suggests that in Au+Au collisions at RHIC, $J/\psi$'s
are suppressed in a medium unlike the medium produced in S+U/Pb+Pb
collisions at SPS energy or in Cu+Cu collisions at RHIC energy.
We also apply the model to explain the preliminary PHENIX data on centrality dependence of $p_T$ broadening for $J/\psi$'s.
Within errors, $p_T$ broadening at RHIC seems to be consistent with that at SPS energy. 

The paper is organised as follows: in section II, we briefly describe  the
QCD based nuclear absorption model. In section III, PHNIX data on the 
centrality dependence 
of $J/\psi$ suppression in  Cu+Cu and in Au+Au collisions are analysed. Centrality
dependence of  $p_T$ broadening of $J/\psi$'s are analysed in
section III.
 Summary and conclusions are drawn in section IV.

\section{QCD based nuclear absorption model}

In the QCD based nuclear absorption model \cite{Chaudhuri:2001zx,Chaudhuri:2003zs,Chaudhuri:2002uf,
Chaudhuri:2006zg,Qiu:1998rz},
$J/\psi$  production  is  assumed  to  be a two step process, (a)
formation of a $c\bar{c}$ pair, which is accurately calculable in
QCD and (b) formation of a $J/\psi$  meson  from  the  $c\bar{c}$
pair,   a   non-perturbative   process,   which  is  conveniently
parameterized. The $J/\psi$ cross section in $pp$ collisions,  at
center of mass energy $\sqrt{s}$ is written as,

\begin{eqnarray} \label{eq1}
\sigma^{J/\psi}_{NN} (s) &&
=K \sum_{a,b} \int dq^2 \left( \frac{\hat \sigma_{ab \rightarrow
cc}} {Q^2} \right) \int dx_F \phi_{a/A}(x_a,Q^2)  \nonumber \\
&&   \phi_{b/B}(x_b,Q^2)   \frac{x_a   x_b}{x_a   +  x_b}  \times
F_{c\bar{c} \rightarrow J/\psi} (q^2), \end{eqnarray}

\noindent  where  $\sum_{a,b}$  runs over all parton flavors, and
$Q^2 = q^2 +4 m_c^2$. The  $K$  factor  takes  into  account  the
higher  order corrections. 
We  have  used the CTEQ5L  parton
distribution function for $\phi(x,Q^2)$ \cite{Lai:1994bb}.
The incoming parton momentum fractions
are fixed by kinematics and are; $x_a
=(\sqrt{x^2_F+4Q^2/s}+x_F)/2$               and              $x_b
=(\sqrt{x^2_F+4Q^2/s}-x_F)/2$.
$\hat \sigma_{ab \rightarrow c\bar{c}}$ are the sub process cross
sections  and are given in \cite{Benesh:1994du}. $F_{c \bar{c} \rightarrow
J/\psi}(q^2)$ is the transition  probability  that  a  $c\bar{c}$
pair  with  relative momentum square $q^2$ evolve into a physical
$J/\psi$ meson. It is parameterized as,

\begin{eqnarray} \label{eq2} F_{c \bar{c} \rightarrow J/\psi} (q^2)
=  && N_{J/\psi} \theta(q^2) \theta({4m^\prime}^2 - 4 m_c^2 -q^2) \nonumber\\
  && \times  (1  -  \frac{q^2}{{4m^\prime}^2  -  4  m_c^2  })
\end{eqnarray}

%%%%%%%%%%%%%%%%%%%%%%%%%%%%%%%%%%%%%%%%%%%%%%%%%%%%%%%%%%%%%%%%%%%%%%
%
\begin{figure}
\centerline{\psfig{figure=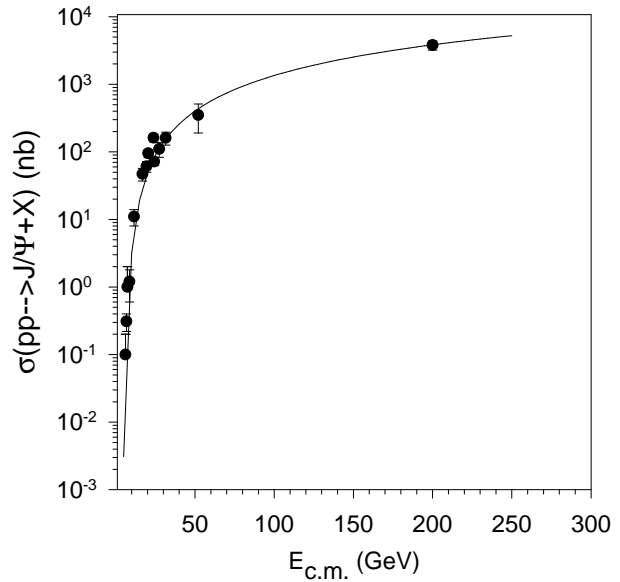,height=13cm,width=9cm}}
\vspace{-4.5cm}  \caption{Energy dependence of total $J/\psi$
cross-section in  pp  collisions.  The  solid  line  is  the  fit
obtained to the data with Eq.1.} \label{F1} \end{figure}
%
%%%%%%%%%%%%%%%%%%%%%%%%%%%%%%%%%%%%%%%%%%%%%%%%%%%%%%%%%%%%%%%%%%%%%%

All  the energy dependence of $J/\psi$ production is contained in
Eq.\ref{eq1}. As shown in Fig.\ref{F1},  with $KN_{J/\psi}$ as a
overall normalisation,
over a wide range of energy, including the RHIC energy,
 the model correctly  reproduces the experimental $J/\psi$ cross
sections in pp collisions.

Our  main  interest  here  is  to  test  the    model against the preliminary PHENIX data on the centrality
dependence of $J/\psi$ suppression in Cu+Cu and Au+Au  collisions
at RHIC energy 
%\cite{Adler:2003rc,Nagle:2002ib,Nagle:2005sh,PereiraDaCosta:2005%xz}. 
\cite{Nagle:2005sh,PereiraDaCosta:2005xz}. 
In AA collisions, at impact
parameter  ${\bf  b}$,  number  of  $J/\psi$  mesons  produced is
calculated as,

\begin{equation}   \label{eq3}   N^{J/\psi}_{AA}({\bf b})      =
\sigma^{J/\psi}_{NN}    \int  d^2s  T_A({\bf s})  T_B({\bf
b-s})  S(L({\bf  b,s})),
\end{equation}

\noindent where  $T_{A,B}$  are the nuclear thickness
function,
\begin{equation} T_A({\bf b})=\int dz \rho_A({\bf  b},z), \label{eq4}
\end{equation}

For the density we use the Woods-Saxon form,

\begin{equation}           \rho(r)=\frac{\rho_0}{1+exp((r-R)/a)},
\hspace{1cm}\int d^3r \rho(r)=A \label{eq5}
\end{equation}

%\noindent with $R=1.12A^{1/3}$fm and $a$=0.54 fm.
\noindent with $R$=6.38 (4.45) fm and $a$=0.535(0.54) fm for Au(Cu)
nucleus \cite{de74}.

In  Eq.\ref{eq3}, $S(L)$ is the suppression factor due to passage of
$J/\psi$  through  a  length  $L$  in  nuclear  environment.  
As mentioned earlier, in the QCD based nuclear absorption model, $J/\psi$'s
are suppressed due to gain in relative 4-square momentum of a
$c\bar{c}$ pair. In  a
nucleon-nucleus/nucleus-nucleus     collision,    the    produced
$c\bar{c}$ pairs interact with the  nuclear  medium  before  they
exit.  Interaction  of a
$c\bar{c}$ pair with the nuclear environment increases the square  of
the  relative  momentum between the pair. As a result,
some of the $c\bar{c}$ pairs  can  gain  enough  relative  square
momentum  to  cross  the threshold to become an open charm meson.
Consequently,  the  cross  section  for  $J/\psi$  production  is
reduced  in comparison with nucleon-nucleon cross section. If the
$J/\psi$ meson travel a distance $L$,  $q^2$  in  the  transition
probability is replaced to,

\begin{equation} \label{eq6}
q^2 \rightarrow q^2 +\varepsilon^2 L,
\end{equation}

\noindent
$\varepsilon^2$  being the relative square momentum gain per unit
length. The length $L({\bf b,s})$ that the $J/\psi$ meson
will traverse is obtained as,

\begin{equation} L({\bf b,s})=n({\bf b,s})/2\rho_0 
\label{eq7}
\end{equation}

\noindent where $n(\bf{b,s})$ is the transverse density,
\begin{equation}      \label{den}      n({\bf      b,s})=T_A({\bf
s})[1-e^{-\sigma_{NN}T_B({\bf b-s})}]  +  [A  \leftrightarrow  B]
\label{eq8}
\end{equation}

 $J/\psi$  suppression  in  the  model  is  governed  by  the
parameter $\varepsilon^2$ and  $L$ (see Eq.\ref{eq6}).
The length $L$ is a geometric term. It has weak energy dependence from the energy dependence of $\sigma_{NN}$, the inelastic NN cross-section. Energy dependence of $J/\psi$ suppression will reflect, most on the parameter $\varepsilon^2$. 
NA50  data on $J/\psi$ production in 450 GeV pp/pA collisions and
in 200 GeV pA collisions are well fitted  with  a  common  square
momentum    gain    factor,    $\varepsilon^2$=0.187   $GeV^2/fm$
\cite{Chaudhuri:2003zs}. The model then explains the centrality dependence of
S+U and Pb+Pb collisions at SPS energy.
As   the  model  parameters  are  fixed  to  reproduce  $J/\psi$
production in pA collisions, where deconfined matter  formation
is unlikely,  it  was  concluded  that  at  SPS  energy  S+U/Pb+Pb 
collisions, $J/\psi$'s are absorbed in a nuclear medium \cite{Chaudhuri:2001zx,Chaudhuri:2003zs}.

If at RHIC energy, $J/\psi$'s are suppressed in a medium denser than the medium produced in SPS energy nuclear collisions, $\varepsilon^2$ will increase. In a denser medium, 
the $c\bar{c}$ pair will interact more with the medium and per 
unit length will gain more square momentum.
Parametric value of $\varepsilon^2$ at RHIC energy then can indicate whether or not, a dense medium is produced.
We note that due enhanced energy charmonium production increases at RHIC, but as shown in
Fig.\ref{F1}, that energy dependence is included in the model.

%%%%%%%%%%%%%%%%%%%%%%%%%%%%%%%%%%%%%%%%%%%%%%%%%%%%%%%%%%%%%%%%%%%%%%
%
\begin{figure}[h]
\centerline{\psfig{figure=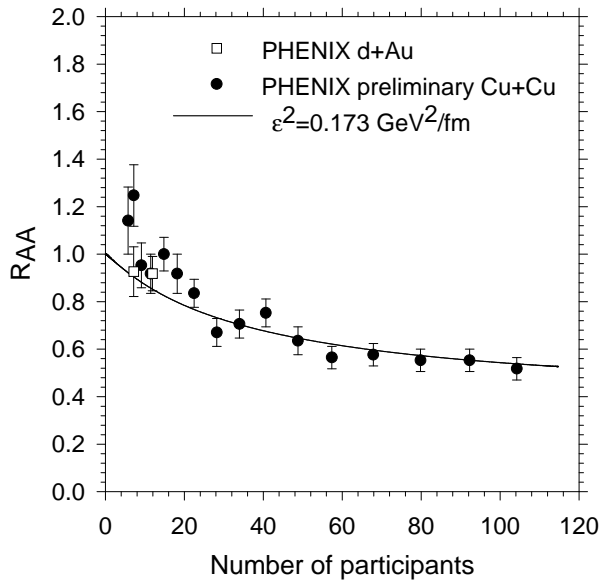,height=13cm,width=9cm}}
\vspace{-4.5cm}   \caption{Preliminary PHENIX data   on   the
centrality dependence of nuclear modification factor for $J/\psi$
in Cu+Cu collisions at RHIC. The solid lines is the best fit in
the  QCD  based  nuclear absorption model with $\varepsilon^2$=0.173 $GeV^2/fm$.} 
\label{F2}
\end{figure}
%
%%%%%%%%%%%%%%%%%%%%%%%%%%%%%%%%%%%%%%%%%%%%%%%%%%%%%%%%%%%%%%%%%%%%%%

\section{$J/\psi$ suppression in Cu+Cu/Au+Au collisions at RHIC}

PHENIX collaboration has measured the centrality dependence of
$J/\psi$ suppression, in Cu+Cu and in Au+Au collisions, in two 
ranges of rapidity intervals, (i) $-0.35 \leq y \leq  0.35$  and 
(ii)$1.2  \leq  y  \leq  2.2$,  \cite{Nagle:2005sh,PereiraDaCosta:2005xz}.
In   Fig.\ref{F2},   preliminary  PHENIX  data  on  the
centrality dependence of nuclear modification  factor  ($R_{AA}$)
for  $J/\psi$,  in Cu+Cu collisions are shown. Two rapidity ranges
of data are not distinguished. We note that the present model  is
designed  for central rapidity only. However, presently we ignore
this limitation of the  model.  As  it  is  evident  from  Fig.\ref{F2},
rapidity dependence of $J/\psi$ suppression is not large in Cu+Cu
collisions.  Using the CERN minimisation programme MINUIT, with $\varepsilon^2$ as a parameter of the QCD based absorption model, we fit the data. The best fit is obtained with
$\varepsilon^2=0.173\pm 0.007$ $GeV^2/fm$. The fit is shown
in Fig.\ref{F2} (the solid line).
With the exception of very peripheral collisions, 
PHENIX data on the centrality dependence of
$R_{AA}$,  for  $J/\psi$  in  Cu+Cu  collisions,
are well explained the QCD  based  nuclear  absorption  model.
In   very peripheral  collisions,  the  model over predicts the suppression.
Indeed, in  very  peripheral  collisions,  PHENIX  data  indicate
enhancement  rather  than suppression of $J/\psi$, presumably due
to Cronin effect. 
%The best fitted value $\varepsilon^2=.173\pm .007$ $GeV^2/fm$
%is marginally less than the value required at SPS energy, %$\varepsilon^2_{SPS}=$0.187 $GeV^2/fm$.

As discussed earlier,  
if   at RHIC energy,  $J/\psi$'s are suppressed in a  
medium, denser than the medium created in SPS energy nuclear collisions,  $\varepsilon^2$ should increase. In Cu+Cu collisions, a contrary result is obtained. 
Compared to $\varepsilon^2$ at SPS energy, at RHIC Cu+Cu 
collisions,
$\varepsilon^2$ decreases by a 
modest 7\%. 
$J/\psi$'s are less suppressed. The result is
consistent with the PHENIX measurement of $J/\psi$ suppression
in d+Au collisions at RHIC \cite{Adler:2005ph}. At RHIC
d+Au collisions, $J/\psi$-nucleon
absorption cross-section, $\sigma_{J/\psi N} \approx $1-3 mb,
is less than the $J/\psi$-nucleon absorption cross-section at
SPS energy, $\sigma_{J/\psi N} \approx $4-5 mb \cite{Cortese:2003iz}. Good fit to the centrality dependence of
$J/\psi$ suppression in Cu+Cu collisions, data, in the QCD based nuclear 
absorption model, with $\varepsilon^2$ close to the value at SPS energy, indicate that in Cu+Cu collisions, $J/\psi$'s are
suppressed in a nuclear medium, much like the medium produced 
in nuclear collisions at SPS energy.

%%%%%%%%%%%%%%%%%%%%%%%%%%%%%%%%%%%%%
%%%%%%%%%%%%%%%%%%%%%%%%%%%%%%%%%%%%%
%
\begin{figure}[h]
\centerline{\psfig{figure=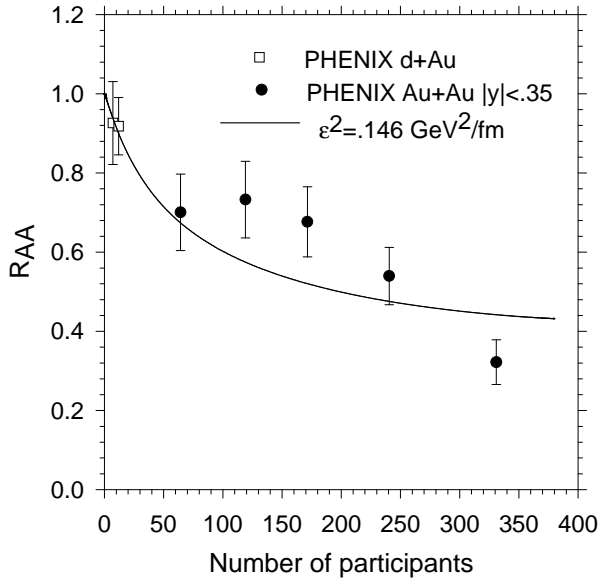,height=13cm,width=9cm}}
\vspace{-4.5cm}   \caption{Preliminary PHENIX  data  on  the
centrality dependence of nuclear modification factor for $J/\psi$
in Au+Au collisions at RHIC. The solid line  is the best fit in the
QCD  based  nuclear absorption model 
with $\varepsilon^2$=0.146 $GeV^2/fm$.  }
\label{F3}
\end{figure}
%
%%%%%%%%%%%%%%%%%%%%%%%%%%%%%%%%%%%%%
%%%%%%%%%%%%%%%%%%%%%%%%%%%%%%%%%%%%%

Next we fit the PHENIX data on $J/\psi$ suppression in Au+Au collisions.
Centrality dependence of  the nuclear modification factor ($R_{AA}$) for $J/\psi$ in
the central rapidity region $-0.35 < y < 0.35$, is  shown  in  Fig.\ref{F3}.  Data points are few and error bars are also large.
The best fit to the data is obtained with 
$\varepsilon^2=0.146\pm .014$ $GeV^2/fm$. It is shown as the solid
line in Fig.\ref{F3}. 
The fit to the data  is not satisfactory. 
In very central collisions,
the model over predict $R_{AA}$ and in mid central collisions the model under predict $R_{AA}$. 
The best fitted value of $\varepsilon^2=0.146 \pm .014$
$GeV^2/fm$ is 
$\sim$ 15\% lower than the value required to explain the
PHENIX data on $J/\psi$ suppression in Cu+Cu collisions.
 Apparently, in Au+Au collisions,
$J/\psi$'s are suppressed in a medium,
less dense than that produced in Cu+Cu collisions. This is inconsistent with other results at RHIC Au+Au collisions 
\cite{BRAHMSwhitepaper,PHOBOSwhitepaper,PHENIXwhitepaper,STARwhitepaper}.  
As mentioned earlier, in RHIC Au+Au collisions, 
deconfined matter can be formed. $J/\psi$'s can be suppressed in the deconfined matter.  QCD based nuclear absorption model  do not account 
for such a suppression. Unsatisfactory
fit to the Au+Au data  may be due to neglect of  deconfined medium production in Au+Au collisions.

%We believe that the present result 
%$\varepsilon^2(Cu+Cu) > \varepsilon^2(Au+Au)$ is an indirect 
%confirmation that a high density, color opaque
%medium of deconfined quarks and gluons formed. 
%In a deconfining
%medium charmonium are suppressed due to screening of color %forces, unlike the mechanism assumed in the
%QCD based nuclear absorption model. We also note that
%in a deconfining medium
%$J/\psi$'s can be regenerated
%\cite{ Thews:2000rj,Braun-Munzinger:2000px},   effectively %reducing the suppression.
%Then if in Au+Au
%collisions, $J/\psi$ are suppressed in a deconfining medium, 
%present model will not be applicable. If applied, it will not give
%consistent results, as we obtain presently. 

%%%%%%%%%%%%%%%%%%%%%%%%%%%%%%%%%%%%%
%%%%%%%%%%%%%%%%%%%%%%%%%%%%%%%%%%%%%
%
\begin{figure}[h]
\centerline{\psfig{figure=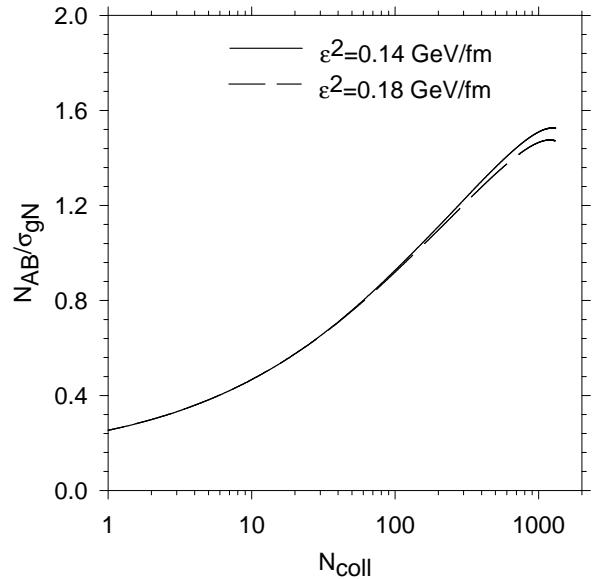,height=13cm,width=9cm}}
\vspace{-4.5cm}   \caption{Collision number dependence of
the ratio of average number of gluon-nucleon collisions
to the gluon-nucleon cross-section in Au+Au collisions. The solid and dashed lines are obtained with $\varepsilon^2$=0.14 and 0.18 $GeV^2/fm$ respectively.}
\label{F4}
\end{figure}

\section{$p_T$ broadening of $J/\psi$ in Cu+Cu and Au+Au collisions}

It  is  well  known  that  in pA and AA collisions, the secondary
hadrons     generally     shows      a      $p_T$      broadening
\cite{ptbr,Kharzeev:1997ry}. $p_T$ broadening of $J/\psi$'s, in
S+U and Pb+Pb collisions are well explained in the QCD based nuclear absorption model. Recently PHENIX collaboration has
measured $p_T$ broadening of $J/\psi$ in Cu+Cu and Au+Au
collisions \cite{Nagle:2005sh,PereiraDaCosta:2005xz}. It is interesting to compare the QCD based nuclear absorption model
predictions with the PHENIX data. 

The  natural  basis for the $p_T$ broadening is the initial state
parton  scatterings.  For  $J/\psi$'s,  gluon  fusion  being  the
dominant  mechanism  for  $c\bar{c}$  production,  initial  state
scattering   of   the   projectile/target   gluons    with    the
target/projectile   nucleons   causes   the   intrinsic  momentum
broadening of  the  gluons,  which  is  reflected  in  the  $p_T$
distribution  of  the  resulting  $J/\psi$'s.  Parameterising the
intrinsic transverse momentum of a gluon, inside a nucleon as,

\begin{equation} f(q_T) \sim exp(-q^2_T/<q^2_T>) \end{equation}

\noindent  momentum  distribution of the resulting $J/\psi$ in NN
collision is obtained by convoluting two such distributions,

\begin{equation}             f^{J/\psi}_{NN}(p_T)            \sim
exp(-p^2_T/<p^2_T>^{J/\psi}_{NN}) \end{equation}

\noindent  where  $<p^2_T>^{J/\psi}_{NN}  =  <q^2_T>+<q^2_T>$. In
nucleus-nucleus collisions at  impact  parameter  ${\bf  b}$,  if
before  fusion, a gluon undergo random walk and suffer $N$ number
of subcollisions, its square momentum  will  increase  to  $q^2_T
\rightarrow  q^2_T  +  N\delta_0$,  $\delta_0$  being the average
broadening in each subcollisions.  Square  momentum  of  $J/\psi$
then easily obtained as,

\begin{equation}     \label{eq11}     <p^2_T>^{J/\psi}_{AB}(b)     =
<p^2_T>^{J/\psi}_{NN} + \delta_0 N_{AB}({\bf b}) \end{equation}

\noindent  where $N_{AB}({\bf b})$ is the number of subcollisions
suffered by the projectile and target gluons with the target  and
projectile nucleons respectively.

Average number of collisions $N_{AB}({\bf b})$ can be obtained in
a Glauber model \cite{Kharzeev:1997ry}. At impact parameter ${\bf
b}$,  the  positions  $({\bf  s},z)$  and  $({\bf b-s},z^\prime)$
specifies the formation point of $c\bar{c}$ in  the  two  nuclei,
with ${\bf s}$ in the transverse plane and $z,z^\prime$ along the
beam  axis.  The  number  of collisions, prior to $c\bar{c}$ pair
formation, can be written as,

\begin{eqnarray}  \label{2}  N(b,s,z,z^\prime)  =  && \sigma_{gN}
\int_{-\infty}^z dz_A \rho_A(s,z_A) \\ \nonumber && + \sigma_{gN}
\int_{-\infty}^{z^\prime}        dz_B        \rho_B(b-s,z^\prime)
\end{eqnarray}

\noindent where $\sigma_{gN}$ is the gluon-nucleon cross-section.
Above  expression  should  be  averaged  over  all  positions  of
$c\bar{c}$ formation with  a  weight  given  by  the  product  of
nuclear densities and survival probabilities $S$,

\begin{eqnarray}\label{3}     &&N_{AB}(b)=     \int     d^2     s
\int^\infty_{-\infty}   dz   \rho_A(s,z)    \int^\infty_{-\infty}
dz^\prime    \rho_B(b-s,z^\prime)    \times    \nonumber  \\
 &&
S({\bf b,s})     N(b,s,z,z^\prime)     /      \int      d^2s
%S(b,s,z,z^\prime)     N(b,s,z,z^\prime)     /      \int      d^2s
\int^\infty_{-\infty}  dz  \rho_A(s,z)  \times    \nonumber \\ &&
\int^\infty_{-\infty}       dz^\prime        \rho_B(b-s,z^\prime)
S(b,s,z,z^\prime) \end{eqnarray}

In Fig.\ref{F4}, the centrality dependence of the ratio $N_{AB}/\sigma_{gN}$, in Au+Au collisions are shown. The
solid and dashed lines corresponds to  $\varepsilon^2$ =0.14 GeV/fm   and 0.18 GeV/fm respectively. $N_{AB}/\sigma_{gN}$ do not
show large dependence on $\varepsilon^2$. Even though 
$\varepsilon^2$ differ by 25\%, $N_{AB}$ differ by less than 3\% 
in central collisions.  In less central collisions, the difference is even less. $p_T$ broadening of $J/\psi$ will
not depend much on the exact value of $\varepsilon^2$. 

$p_T$  broadening  of  $J/\psi$'s in AA collisions depends on two
parameters,  (i)  $<p^2_T>^{J/\psi}_{NN}$,  the  mean  squared  transverse
momentum  in  NN  collisions and (ii) the
product of the gluon-nucleon cross-section and the average parton
momentum broadening per collision,  $\sigma_{gN}\delta_0$.  
$<p^2_T>^{J/\psi}_{NN}$ is a measured in RHIC energy
p+p collisions, $<p^2_T>^{J/\psi}_{NN} = 4.2 \pm 0.7$ $GeV^{2}$. The other parameter, $\sigma_{gN}\delta_0$ is
essentially non-measurable, as gluons are not free. Its value can only be obtained from experimental data on  $p_T$ broadening 
of $J/\psi$. At SPS energy S+U/Pb+Pb collisions $\sigma_{gN}\delta_0$ is estimated as $0.442 \pm 0.056$ $GeV^2$ \cite{Chaudhuri:2006zg}. PHENIX data on $J/\psi$ $p_T$ broadening can be used to estimate its value at RHIC energy.

\begin{figure}[h]
\centerline{\psfig{figure=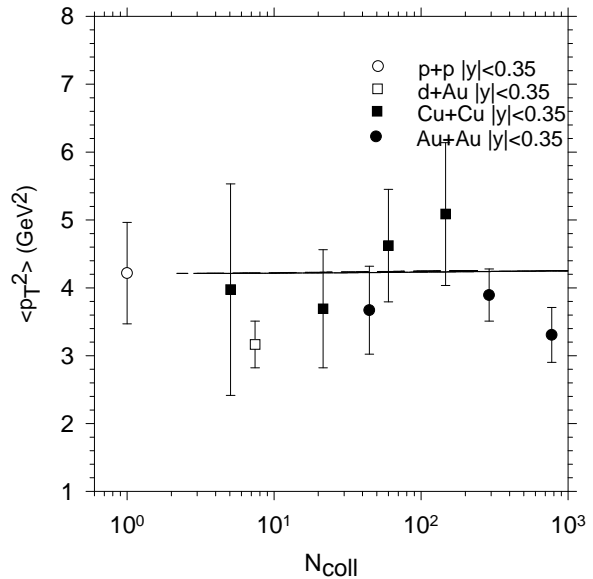,height=13cm,width=9cm}}
\vspace{-4.5cm}   \caption{$J/\psi$ mean square transverse momentum as a function of collision number, in p+p, d+Au, Cu+Cu and Au+Au collisions are shown. The solid and
dashed lines are fit to the Au+Au and Cu+Cu data respectively.} 
\label{F5}
\end{figure}

In Fig.\ref{F5}, PHENIX data on the centrality dependence of mean square transverse momentum $<p^2_T>$, in Cu+Cu and in Au+Au collisions are shown. For comparison purpose, $<p^2_T>$ in
in p+p and in d+Au collisions are also shown.
  As data points are few, we do not fit the individual Cu+Cu or Au+Au
data sets. Rather we fit the combined data sets.  
We fix  $<p_T^2>_{NN}$ at the measured central value,
$<p^2_T>_{NN}$ = 4.2 $GeV^2$, and vary $\sigma_{gN}\delta_0$. Best fit is obtained with  $\sigma_{gN}\delta_0 =0.03 \pm 0.51$ $GeV^2$. 
RHIC data do not show any evidence of $p_T$ broadening,
as indicated by very small value of $\sigma_{gN}\delta_0$.
In Fig.\ref{F5}, model predictions with the central value, $\sigma_{gN}\delta_0$ =0.03 $GeV^2$  are shown. The solid
and dashed  lines are for Au+Au and Cu+Cu collisions respectively. The predictions for Cu+Cu collisions closely match
that for Au+Au collisions and cannot be distinguished. 

%However, poor quality of the data, 
%present data do not allow accurate estimate of
%$\sigma_{gN}\delta_0$ at RHIC energy.  

%%%%%%%%%%%%%%%%%%%%%%%%%%%%%%%%%%%%%
%%%%%%%%%%%%%%%%%%%%%%%%%%%%%%%%%%%%%
%
%
%%%%%%%%%%%%%%%%%%%%%%%%%%%%%%%%%%%%%
%%%%%%%%%%%%%%%%%%%%%%%%%%%%%%%%%%%%%
 
\section{Summary and conclusions}

To  summarize, we have analysed the (preliminary) PHENIX 
data \cite{Adler:2003rc,Nagle:2002ib} on the centrality dependence of $J/\psi$ suppression and $p_T$ broadening, in
Cu+Cu and in Au+Au collisions at RHIC energy, $\sqrt{s}$=200 
GeV. The data are analysed in the  QCD  based  nuclear  absorption  model
\cite{Chaudhuri:2001zx,Chaudhuri:2003zs,Chaudhuri:2002uf,
Chaudhuri:2006zg,Qiu:1998rz}.
In the  model, $J/\psi$'s
suppression is controlled by a parameter $\varepsilon^2$.  Larger the $\varepsilon^2$, more is
the suppression. Centrality dependence of $J/\psi$ suppression at SPS energy require $\varepsilon^2_{SPS}$=0.187 $GeV^2/fm$.
With $\varepsilon^2$ as a parameter, we have fitted the
PHENIX data on the
centrality dependence of $J/\psi$ suppression in Cu+Cu and in Au+Au collisions. Cu+Cu data 
are well explained  with $\varepsilon^2 =0.173 \pm 0.007$ $GeV^2/fm$, close to the SPS energy value. In Cu+Cu collisions,
$J/\psi$'s are suppressed in a medium much like the medium created in SPS energy collisions. 
No exotic, high-density matter is created in Cu+Cu collisions.

Centrality dependence of $J/\psi$ suppression, in Au+Au collisions, is not well explained in the model. Best fitted value,  $\varepsilon^2=0.146 \pm .014$ $GeV^2/fm$, over predict the suppression in
mid-central collisions and under predict the suppression in very central collisions. We conclude that in Au+Au collisions,
$J/\psi$'s are suppressed in medium unlike the medium
created in SPS energy nuclear collisions or in Cu+Cu collisions at RHIC energy. 
We have also analysed the PHENIX data on 
$p_T$ broadening of $J/\psi$ in Cu+Cu and Au+Au collisions.
RHIC data on $p_T$ broadening do not show any evidence of
$p_T$ broadening.
%Poor quality of the data do not allow any definitive conclusion.
%It can only be said that within errors, $p_T$ broadening at RHIC %is consistent with $p_T$ broadening at SPS energy.

\end{document}